\documentclass[english,twocolumn,prb,superscriptaddress]{revtex4-2}
\usepackage{amssymb}
\usepackage{amsmath}
\usepackage{bbm}
\usepackage{epsfig}
\usepackage{color}
\usepackage{multirow}
\usepackage{booktabs}
\usepackage{dsfont}
\usepackage{tikz}
\usepackage{appendix}
\usepackage[colorlinks=true,linkcolor=blue,citecolor=blue,urlcolor=blue]{hyperref}
\newcommand{\ket}[1]{\left| #1 \right>} 
\newcommand{\bra}[1]{\left< #1 \right|} 

\begin{document}

\title{Generic reduction theory for Fermi sea topology in metallic systems}

\author{Wei Jia}
\email{jiaw@lzu.edu.cn}
\affiliation{Lanzhou Center for Theoretical Physics, Key Laboratory of Theoretical Physics of Gansu Province, Key Laboratory of Quantum Theory and Applications of MoE, Gansu Provincial Research Center for Basic Disciplines of Quantum Physics, Lanzhou University, Lanzhou 730000, China}

\begin{abstract}
The Fermi sea of a metal can host exotic quantum topology, which governs its conductance quantization and is characterized by the Euler characteristic ($\chi_F$). In contrast to the well-known band topology, which is determined by the global features of wave functions, the topology of such metallic systems is intrinsically linked to the geometry of the Fermi sea. As a result, probing and identifying $\chi_F$ in high-dimensional systems presents a challenge. Here, we propose a generic dimensional reduction theory for the Fermi sea topology in $d$-dimensional metallic systems, showing that $\chi_F$ can be determined by the features of so-called reduced critical points on Fermi surfaces. Moreover, we reveal that $\chi_F$ can be interpreted as a topological invariant of band topology by mapping a metallic system to a gapped system. Building on this nontrivial result, we identify a broad class of topological superconductors (SCs) whose topological numbers are precisely determined by the $\chi_F$ of their normally filled bands. This provides an indirect method to capture $\chi_F$ by measuring the (pseudo)spin polarizations of these topological SCs. Our findings are expected to significantly advance research into Fermi sea topology.
\end{abstract}

\maketitle
\section{Introduction}

Since the discovery of the two-dimensional (2D) integer quantum Hall effect~\cite{klitzing1980new}, quantum topology has played an important role in the field of condensed matter physics. One of the most fundamental phenomena in topological quantum phases is quantized response~\cite{thouless1982quantized}, which arises from the global features of wave functions across the Brillouin zone (BZ) and is characterized by topological invariants of the ground state~\cite{hasan2010colloquium,qi2011topological}. In metallic systems, another form of quantum topology emerges, significantly influencing the quantized response~\cite{landauer1957spatial,fisher1981relation,buttiker1986four,stone1988measured}. This topology is governed by the geometry of the Fermi sea rather than the wave functions. Although the associated conductance quantization is less robust than the Hall conductance, it has been experimentally observed in various systems, including quantum point contacts~\cite{van1988quantized}, semiconductor nanowires~\cite{honda1995quantized,van2013quantized}, and carbon nanotubes~\cite{frank1998carbon}. These observations raise a fundamental question: how can the topology of the Fermi sea be precisely characterized? 

Recently, a breakthrough has demonstrated that the Fermi sea topology is characterized by the Euler characteristic $\chi_F$~\cite{kane2022quantized}, which can be identified through critical points where the Fermi velocity vanishes within the filled bands. Several theoretical probing schemes for $\chi_F$ have been proposed in $2$D systems, such as multipartite entanglement~\cite{tam2022topological}, Andreev state transport~\cite{tam2023probing,tam2023topological}, and density correlations of Fermi gases~\cite{tam2024topological}. Additionally, the quantized response has been predicted to be measurable in ultracold atomic gases~\cite{yang2022quantized,zhang2023quantized}. However, the observation of $\chi_F$ remains challenging in $d$D metallic systems. This difficulty arises because $\chi_F$ is associated with the properties of filled bands, where the critical points are not easily resolved in high-dimensional systems. There is an urgent need to develop a simpler characterization of $\chi_F$ to facilitate its detection in generic metallic systems. Furthermore, electrons on Fermi surfaces (FSs) can interact to produce Cooper pairs, leading to the emergence of topological superconductors (SCs)~\cite{sato2017topological}. The number of Majorana edge states~\cite{fu2008superconducting,wilczek2009majorana} in these topological SCs is closely linked to the Fermi sea topology of their normally filled bands~\cite{poon2018semimetal,jia2019topological,yang2023euler}. Therefore, uncovering this subtle and nontrivial connection is essential, as it may pave the way for discovering other novel physical effects stemming from the topology of the Fermi sea.

In this paper, we propose a generic dimensional reduction theory for the $d$D Fermi sea topology, showing that  the Euler characteristic $\chi_F$ can be determined by specific discrete momentum points on the $(d-1)$D FSs, referred to as reduced critical points. Unlike the original critical points, which require all components of the Fermi velocity to vanish for characterizing $\chi_F$~\cite{kane2022quantized}, the reduced critical points are defined by the vanishing of only one component of the Fermi velocity, greatly simplifying the calculation of $\chi_F$. Moreover, we reveal that $\chi_F$ can be interpreted as a topological invariant of the band topology by mapping a single metallic band to two gapped bands. Specifically, for the topological phases classified by $\mathbb{Z}$, $2\mathbb{Z}$, and $\mathbb{Z}_2$, which are characterized by the topological invariants $\mathcal{W}$, $2\mathcal{W}$, and $(-1)^{\mathcal{W}}$ respectively, this mapping precisely yields $\mathcal{W}=\chi_F$. Additionally, when the metallic systems are driven into topological SCs through a nonzero pairing order parameter induced by the Hubbard interaction among electrons on the FSs, we further establish that the topological number of these SCs is directly determined by the $\chi_F$ of their normally filled bands. This provides an indirect method to detect $\chi_F$ by measuring the (pseudo)spin polarizations in the topological SCs. Our findings not only simplify the characterization of Fermi sea topology but also establish a profound connection between Fermi sea geometry and topological SCs. 

The remaining part of this paper is organized as follows. In Sec.~\ref{Generic characterization}, we present a generic framework for characterizing Fermi sea topology. In Sec.~\ref{Mapping}, we introduce a nontrivial mapping from a single metallic band to two gapped bands. In Sec.~\ref{Reduce}, we develop a dimensional reduction approach to calculate the Euler characteristic efficiently. In Sec.~\ref{SC}, we explore the Fermi sea topology for the normally filled bands of topological SCs. In Sec.~\ref{measure}, we propose a theoretical scheme to measure the Fermi sea topology through the (pseudo)spin polarizations. Finally, we provide a brief discussion and summary in Sec.~\ref{discussion}.

\section{Generic characterization of Fermi sea topology}\label{Generic characterization}

Our starting point is a $d$D metallic system with the electronic band $E_\mathbf{k}$ in momentum space $\mathbf{k}$. The topology of the Fermi sea in this system governs the quantization of conductance and is characterized by the Euler characteristic $\chi_F$~\cite{kane2022quantized}. Building on Morse theory~\cite{milnor1963morse,nash1988topology}, we present a generic method to calculate $\chi_F$, expressed as
\begin{equation}
\chi_F=\sum_m\eta_m.
\label{eq:Euler_index}
\end{equation} 
Here $m$ labels the critical points $\mathbf{k}_m$ in $E_\mathbf{k}$ at which the Fermi velocity $\mathbf{v}_\mathbf{k}=\nabla_\mathbf{k}E_{\mathbf{k}}$ vanishes, i.e., $\mathbf{v}_\mathbf{k}=0$, for $E_\mathbf{k}<E_F$, with $E_F$ denoting the Fermi energy. For simplicity, we hereby set $E_F=0$. The index $\eta_m$ characterizes the topological signature of the $m$th critical point and is given by
\begin{equation}
\eta_m=\frac{\Gamma(d/2)}{2\pi^{d/2}}\int_{\mathcal{L}_m}\frac{1}{|\mathbf{v}_{\mathbf{k}}|^d}\sum^d_{i=1}(-1)^{i-1}v_{i,\mathbf{k}}\bar{\bigwedge}^{d}_{s=1}\text{d}v_{s,\mathbf{k}},
\label{etm}
\end{equation}
where $\Gamma(a)$ is $\Gamma$ function. Here we define
$
\bar{\bigwedge}^{d}_{s=1}\text{d}v_{s,\mathbf{k}}\equiv \text{d}v_{1,\mathbf{k}}\wedge\cdots\wedge\widehat{\text{d}v_{i,\mathbf{k}}}\wedge\cdots\wedge\text{d}v_{d,\mathbf{k}},
$
where the term $\text{d}v_{i,\mathbf{k}}$ (marked by a big hat) is omitted. The contour ${\cal L}_m$ denotes a $(d-1)$D surface enclosing the critical point $\mathbf{k}_m$. It is clear that Eq.~\eqref{etm} describes a mapping from a $(d-1)$D torus to a $(d-1)$D sphere, i.e., $\mathsf{T}^{(d-1)}\mapsto\mathsf{S}^{(d-1)}$, indicating that $\eta_m$ is an integer-valued topological index. 

Generally, the Fermi velocity near a critical point $\mathbf{k}_m$ can exhibit two distinct behaviors: nonlinear dispersion or linear dispersion. At $\mathbf{k}_m$, the former corresponds to a degenerate critical point (DCP) and the latter corresponds to a nondegenerate critical point (NDCP). For a DCP, the filled band $E_\mathbf{k}$ satisfies $\text{det}[\mathbb{H}]=0$, where 
\begin{equation}
\mathbb{H}=\begin{bmatrix}
\frac{\partial^2E_\mathbf{k}}{\partial k_1^2} & \frac{\partial^2E_\mathbf{k}}{\partial k_1\partial k_2} & \cdots & \frac{\partial^2E_\mathbf{k}}{\partial k_1\partial k_d}\\
\\
\frac{\partial^2E_\mathbf{k}}{\partial k_2\partial k_1} & \frac{\partial^2E_\mathbf{k}}{\partial k_2^2} & \cdots & \frac{\partial^2E_\mathbf{k}}{\partial k_2\partial k_d}\\
\\
\vdots & \vdots & \ddots & \vdots\\
\\
\frac{\partial^2E_\mathbf{k}}{\partial k_d\partial k_1} & \frac{\partial^2E_\mathbf{k}}{\partial k_d\partial k_2} & \cdots & \frac{\partial^2E_\mathbf{k}}{\partial k_d^2}\\
\end{bmatrix}
\end{equation}
is the Hessian matrix of $E_\mathbf{k}$. Consequently, each DCP is associated with a high integer-value $\eta_m$. In contrast, an NDCP is characterized by $\text{det}[\mathbb{H}]\neq 0$, which simplifies Eq.~\eqref{etm} to $\eta_m=\text{sgn}\{\text{det}[\mathbb{H}]\}$. This implies that each NDCP contributes a unit value of $\eta_m$. When a DCP or NDCP passes though $E_F$, it allows $\chi_F$ to change at a Lifshitz transition~\cite{lifshitz1960anomalies}. Notably, the nondegenerate case has been discussed in Ref.~\cite{kane2022quantized}, and we have proven in Appendix~\ref{appendix-1} that it is a special case of Eq.~\eqref{etm}.  

\begin{figure}[t]
\centering
\includegraphics[width=1.0\columnwidth]{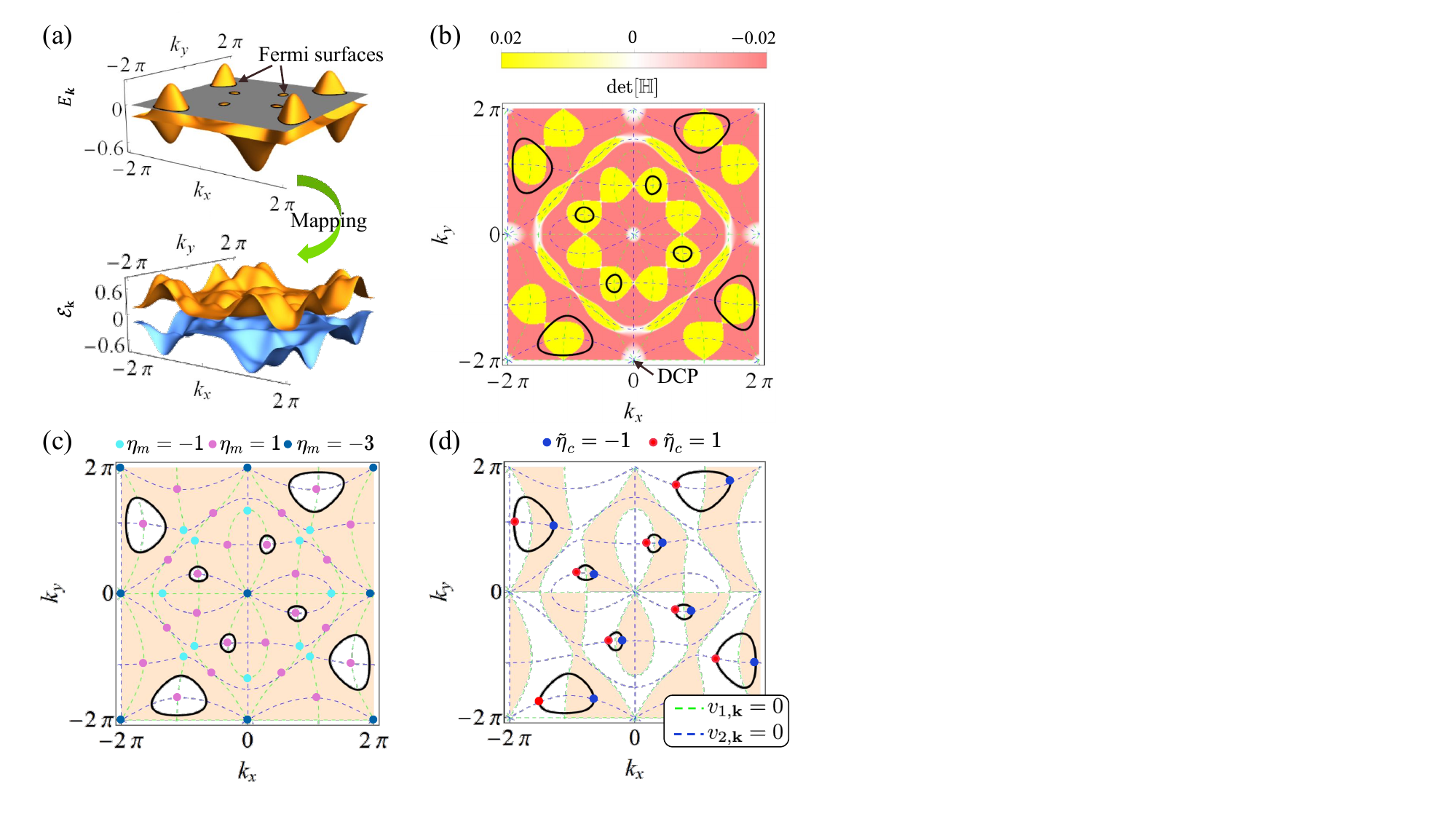}
\caption{(a) A 2D filled band $E_\mathbf{k}$ with eight hole-like FSs (black curves), which can be translated to two gapped energy bands $\mathcal{E}_{\mathbf{k}}$ by the mapping. (b) Numerical results of $\text{det}[\mathbb{H}]$, where both $v_{1,\mathbf{k}}=0$ (green dashed lines) and $v_{2,\mathbf{k}}=0$ (blue dashed lines) determine 40 critical points. Only four critical points at $(k_x,k_y)=(0,0)$, $(0,-2\pi)$, $(-2\pi,0)$, and $(-2\pi,-2\pi)$ are DCPs and the others are NDCPs. (c) Signature of each critical point, characterized by $\eta_m$. The Fermi sea topology is captured by $\mathbf{k}_m$ in the region of $E_\mathbf{k}<0$, giving $\chi_F=\sum_m\eta_m=-8$. (d) Positive (red points) or negative (blue points) reduced critical points, located on the 1D FSs, where $\tilde{k}_c$ in the region of $v_{1,\mathbf{k}_c}<0$ gives $\chi_F=\sum_c\tilde{\eta}_c=-8$.}
\label{fig:1}
\end{figure}

The above characterization can be illustrated by using a simple $2$D filled band $E_\mathbf{k}=e_{1,\mathbf{k}}e_{2,\mathbf{k}}e_{3,\mathbf{k}}-\mu$, where
\begin{equation}
\begin{split}
&e_{1,\mathbf{k}}=\sin^2\left(\frac{k_x}{2}\right)-\sin^2\left(\frac{k_y}{2}\right)\\
&e_{2,\mathbf{k}}=\sin\left(\frac{k_x}{2}\right)\sin\left(\frac{k_y}{2}\right)\\
&e_{3,\mathbf{k}}=m_z-t_s\left[\cos\left(\frac{k_x}{2}\right)+\cos\left(\frac{k_y}{2}\right)\right].
\end{split}
\end{equation}
Here the BZ is $k_{x/y}\in[-2\pi,2\pi)$, with $k_{x,y}=k_{1,2}$. The parameters $m_z$, $\mu$, and $t_s$ are constants, where $\mu$ serves to adjust the Fermi energy. Setting $m_z=0.5t_s$ and $\mu=0.18t_s$, we observe that $E_\mathbf{k}$ is partially filled and exhibits eight hole-like FSs [see Fig.~\ref{fig:1}(a)]. Meanwhile, the conditions $v_{1,\mathbf{k}}=0$ and $v_{2,\mathbf{k}}=0$ yield $40$ critical points in the BZ, of which four critical points at $(k_x,k_y)=(0,0)$, $(0,-2\pi)$, $(-2\pi,0)$, and $(-2\pi,-2\pi)$ are degenerate [see Fig.~\ref{fig:1}(b)]. Using Eq.~\eqref{etm}, we find these DCPs are characterized by $\eta_m=-3$. The remaining critical points are non-degenerate, with $\eta_m=\pm 1$. Since there are $28$ NDCPs and four DCPs in the filled band [see Fig.~\ref{fig:1}(c)], the Fermi sea topology is determined by $\chi_F=-8$. Furthermore, when the band is completely filled, all the critical points are within Fermi sea, resulting in $\chi_F=0$.

\section{Mapping from metallic bands to gapped bands}
\label{Mapping}

The Fermi sea topology for a single metallic band can be connected to the band topology by mapping this metallic band to two gapped bands: a conduction band and a valence band. By redefining 
\begin{equation}
E_\mathbf{k}\equiv v_{0,\mathbf{k}},~~~~\mathbf{v}_\mathbf{k}\equiv(v_{1,\mathbf{k}},v_{2,\mathbf{k}},\cdots,v_{d,\mathbf{k}}),
\end{equation}
we construct a $d$D Dirac-type Hamiltonian 
\begin{equation}
H_\mathbf{k}=v_{0,\mathbf{k}}\gamma_0+\sum^d_{i=1}v_{i,\mathbf{k}}\gamma_i.
\label{gapsystem}
\end{equation}
Here the $\boldsymbol{\gamma}$ matrices satisfy the anticommutation relation, ensuring that $H_\mathbf{k}$ hosts two gapped bands with $\mathcal{E}_{\mathbf{k}}=\pm\sqrt{\sum^d_{i=0}v^2_{i,\mathbf{k}}}$, as shown in Fig.~\ref{fig:1}(a). This mapping transforms the single metallic band into two gapped bands, i.e., $E_\mathbf{k}\mapsto \mathcal{E}_\mathbf{k}$. Physically, this mapping can be understood through two key processes. First, it induces a band inversion of $E_{\mathbf{k}}$ at the Fermi level $E_F$, primarily driven by the $v_{0,\mathbf{k}} \gamma_0$ term. This term alone results in two gapless bands, $\pm E_{\mathbf{k}}$. Subsequently, the remaining terms, where $v_{i,\mathbf{k}}$ couples with $\gamma_i$, act as an effective (pseudo)spin-orbit coupling, opening a band gap and yielding the two bands $\mathcal{E}_\mathbf{k}$.

Under the constraint of $E_\mathbf{k}=-E_{-\mathbf{k}}$ or $E_\mathbf{k}=E_{-\mathbf{k}}$, we find that $v_{0,\mathbf{k}}$ has the opposite parity to $v_{i,\mathbf{k}}$. This implies that $H_\mathbf{k}$ can host $\mathbb{Z}$-, $2\mathbb{Z}$-, and $\mathbb{Z}_2$-classified topological phases~\cite{chiu2016classification}, characterized by $\mathcal{W}$, $2\mathcal{W}$, and $(-1)^{\mathcal{W}}$, respectively~\cite{von2021unification}. Specifically, $\mathcal{W}$ is determined by the summation of topological charges in regions where $v_{0,\mathbf{k}}<0$~\cite{zhang2018dynamical,zhang2019dynamical,jia2021dynamically},
\begin{equation}
{\cal W}=\sum_m{\cal C}_m,
\label{eq:Topocharge}
\end{equation} 
where the $m$th topological charge ${\cal C}_m$ is located at $\mathbf{k}_m$ and characterized by the same formula as Eq.~\eqref{etm}, i.e., $\mathcal{C}_m=\eta_m$. Comparing Eq.~\eqref{eq:Euler_index} and Eq.~\eqref{eq:Topocharge}, we observe that $\chi_F$ is exactly equal to $\cal W$; that is 
\begin{equation}
\chi_F=\mathcal{W},
\end{equation}
where $\eta_m$ and FSs in the metallic system are mapped to ${\cal C}_m$ and $v_{0,\mathbf{k}}=0$ in the gapped system, respectively. Thus, $\chi_F$ is interpreted as a topological invariant of the band topology by this mapping. Although the two quantum topologies are fundamentally distinct, the band topology derived from mapping the metallic band offers valuable insights into the Fermi sea topology, including the dimensional reduction of $\chi_F$, the relationship between the band topology of topological SCs and $\chi_F$ of their normal bands, as well as the detection of $\chi_F$ via the (pseudo)spin polarizations of topological SCs. We will present these significant results in the following sections.

\section{Dimensional reduction of Euler characteristic}
\label{Reduce}

In this section, we demonstrate that the $d$D Fermi sea topology can be reduced to a $0$D topology defined by special momentum points in the FSs, simplifying the determination of the Euler characteristic $\chi_F$. To derive the $0$D topological characterization of $\chi_F$, we first revisit the $d$D band topology. It is well-known that a $d$D band topology can be reduced to a 1D winding number, defined by the 1D effective Hamiltonian (see Appendix~\ref{appendix-2}) 
\begin{equation}
H_{\tilde{k}}=v_{d-1,\tilde{k}}\tilde{\gamma}_{d-1}+v_{d,\tilde{k}}\tilde{\gamma}_{d},
\end{equation}
where $\tilde{\boldsymbol{\gamma}}$ denote the effective Gamma matrices, and the $1$D effective BZ is given by $\tilde{k}\equiv\left\{\mathbf{k}|v_{j,\mathbf{k}}=0; j=0,1,\cdots ,d-2\right\}$. As a result, the 1D winding number is further reduced to the total topological charges defined by $H_{\tilde{k}}$ in regions where $v_{d-1,\tilde{k}}<0$ (see Appendix~\ref{appendix-2}),
\begin{equation}
{\cal W}=\sum_c\text{sgn}\left(\frac{\partial v_{d,\tilde{k}}}{\partial \tilde{k}}\right), 
\label{reduce_charge}
\end{equation}
where $\mathcal{C}_c=\text{sgn}({\partial v_{d,\tilde{k}}}/{\partial \tilde{k}})$ defines the $c$th topological charge, located at discrete momentum point $\tilde{k}_c\equiv\{\tilde{k}|v_{{d},\tilde{k}}=0\}$. Based on the previous mapping, we finally obtain
\begin{equation}
\chi_F=\sum_c\frac{(-1)^q}{2}\left[\text{sgn}\left(v_{d,\tilde{k}_{c,\text{R}}}\right)-\text{sgn}\left(v_{d,\tilde{k}_{c,\text{L}}}\right)\right],
\label{Reduce_F}
\end{equation}
where $q=0$ ($1$) corresponds to electron-like (hole-like) $\tilde{k}$. Here, the subscripts R and L denote the right- and left-hand point closest to $\tilde{k}_c$, respectively. Notably, the partial derivative in Eq.~\eqref{reduce_charge} is expressed as a central difference in the $\tilde{k}_c$ domain in Eq.~\eqref{Reduce_F} (see Appendix~\ref{appendix-3}). This reduction maps $\mathbf{k}_m$ in the $d$D system to $\tilde{k}_c$ in the $1$D subsystem, which we refer to as {\it reduced critical point}. These points are located at $\mathbf{k}_c\equiv\left\{\mathbf{k}|v_{j,\mathbf{k}}=0; j=0,1,\cdots,d~\&~j\neq d-1\right\}$ in the original BZ. The signature of each $\tilde{k}_c$ is given by
\begin{equation}
\tilde{\eta}_c=\frac{(-1)^q}{2}[\text{sgn}(v_{d,\tilde{k}_{c,\text{R}}})-\text{sgn}(v_{d,\tilde{k}_{c,\text{L}}})].
\label{rcp}
\end{equation}
To illustrate these results, we consider the $2$D filled band of $E_\mathbf{k}=e_{1,\mathbf{k}}e_{2,\mathbf{k}}e_{3,\mathbf{k}}-\mu$ discussed in Sec.~\ref{Generic characterization}. It is observed that there are eight reduced critical points in the region of $v_{1,\tilde{k}}<0$, determining the Fermi sea topology as $\chi_F=\sum_c\tilde{\eta}_c=-8$ [see in Fig.~\ref{fig:1}(d)]. This approach indeed simplifies the characterization of $\chi_F$.

\begin{figure}[t]
\centering
\includegraphics[width=1.0\columnwidth]{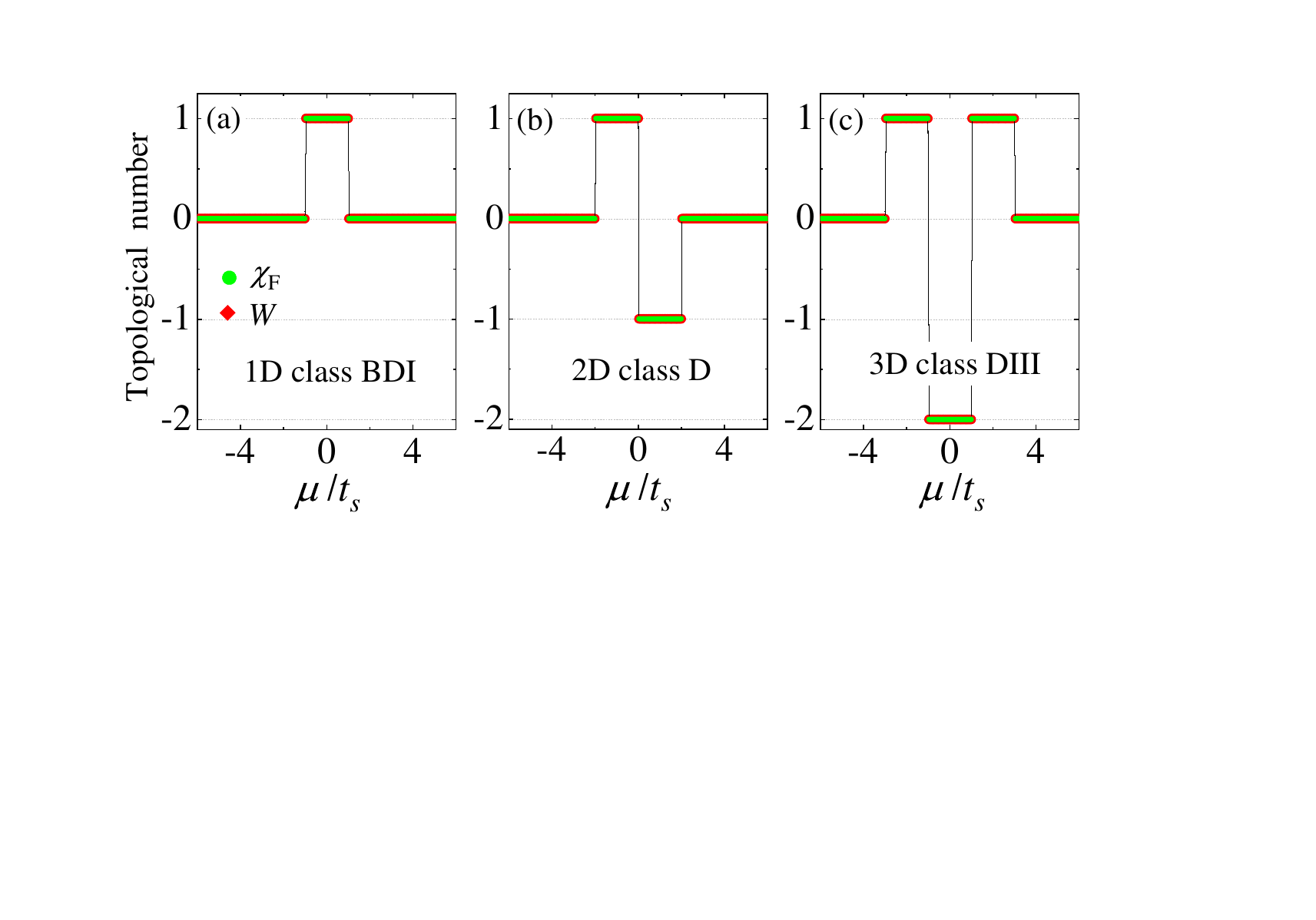}
\caption{Numerical results of $\chi_F$ of the normally filled band and the topological invariant ${\cal W}$ for the 1D Kitaev chain (class BDI) in (a), the 2D $p-\mathtt{i}p$ SC (class D) in (b), and the 3D He-3 B phase (class DIII) in (c).}
\label{fig:2}
\end{figure}

Beside, it is important to emphasize that the reduced critical points provide four key advantages in characterizing $\chi_F$. (i) Determining the features of traditional critical points require all components of the Fermi velocity. In contrast, the features of reduced critical points are determined by one Fermi velocity component [see Eq.~\eqref{rcp}]. (ii) The number of reduced critical points on the FSs is significantly fewer than that of the traditional critical points in the Fermi sea [see Fig.~\ref{fig:1}(c) and \ref{fig:1}(d)],. This simplifies the process of determining $\chi_F$. (iii) In 2D systems, the reduced critical points are linked to the properties of Andreev bound states in metal-superconductor junction~\cite{tam2023probing}, which provides a method to measure $\chi_F$. Remarkably, our theory presents a general expression for identifying $\chi_F$ in $d$D systems, which could improve its experimental detection. (iv) The characterization of $\mathbf{k}_c$ in our theory is different from that in Ref.\textsc{~\cite{tam2023probing}}, which employs the convexity or concavity of FSs. However, our theory introduces $\tilde{\eta}_c$ through the zeroth Chern number of $\mathbf{k}_c$, as demonstrated in 1D FSs [see Fig.~\ref{fig:1}(d)]. This reveals that the reduced critical points are nondegenerate, with their signatures uniquely defined by a unit value, i.e., a fundamental property of the zeroth Chern number (see Appendix~\ref{appendix-4}). Therefore, this simplifies the topology of Fermi sea hosting the DCPs. 

\section{Fermi sea topology of normally filled bands of topological superconductors}
\label{SC}

Mapping the metallic system to the gapped system establishes a nontrivial connection between $\chi_F$ and the band topology. This motivates us to explore the relationship between the Fermi sea topology of the normally filled bands and their corresponding band topologies in the topological SCs. The later, in fact, determines the number of Majorana edge states of the systems~\cite{fu2008superconducting,wilczek2009majorana}. To illustrate this, we consider a special type of the pairing order parameter defined as
\begin{equation}
\Delta_\mathbf{k} \propto \nabla_\mathbf{k}{E}_\mathbf{k}\cdot\mathbf{S},
\label{pairing}
\end{equation}
where the matrices $S_{i}$ have dimension $2^{(d-2)/2}$ [$2^{(d-1)/2}$] for even (odd) $d$, with $i=1,2,\cdots d$. With this paring, the Bogoliubov-de Gennes (BdG) Hamiltonian 
\begin{equation}
\cal H_\mathbf{k}=
\begin{bmatrix}
E_\mathbf{k} & \Delta_\mathbf{k}\\
\Delta^\dagger_\mathbf{k} & -E_\mathbf{k}
\end{bmatrix}
\label{bdgh}
\end{equation}
exhibits a form analogous to the previous gapped Hamiltonian. And then, $\chi_F$ of the normally filled band is exactly equal to the topological invariant ${\cal W}$ of its band topology. Here, the $\mathbb{Z}$-, $2\mathbb{Z}$-, and $\mathbb{Z}_2$-classified topological SCs are characterized by $\mathcal{W}$, $2\mathcal{W}$, and $(-1)^\mathcal{W}$, respectively. In the following, we present the results of the $\mathbb{Z}$-classified topological SCs. The corresponding results for the $2\mathbb{Z}$- and $\mathbb{Z}_2$-classified topological SCs are provided in Appendix ~\ref{appendix-5}.

\begin{table}[t]
	\caption{The $\mathbb{Z}$-classified topological (crystalline) superconductors with the dimensionality $d=4n+j$, where $j=1,2,3,4$ and $n=0,1,2,\cdots$. The BdG Hamiltonian can have time-reversal symmetry $(T)$, particle-hole symmetry $(P)$, chiral symmetry $(C)$, and twofold-rotation symmetry $(R)$. The superscript of $R$ indicates the sign of $R^2$, and the subscript of $R$ specifies the commutation $(+)$/anticommutation $(-)$ relation between $R$ and $T$ and/or $P$. The outside and inside of $(*)$ in the second column shows the topological classes for the even and odd $n$, respectively.} \centering
	\begin{tabular}{c  c  c  c}
		\hline\hline
~~~~~~Dimension~~~&~~~Class~~~&~~~Symmetry~~~&~~$\pi_d$~~~~~~\\ [1ex]
		\hline
~~~$d=4n+1$~~~&~~~~BDI (CII)~~~&~~~${\it T}$,~${\it P}$,~${\it C}$~~~&~$\mathbb{Z}$~~\\[1ex]
~~~$d=4n+2$~~~&~~~~D (C)~~~&~~~${\it P}$~~~&~$\mathbb{Z}$~~\\[1ex]
~~~$d=4n+3$~~~&~~~~DIII (CI)~~~&~~~${\it T}$,~${\it P}$,~${\it C}$~~~&~$\mathbb{Z}$~~\\[1ex]
~~~$d=4n+4$~~~&~~~~C (D)~~~&~~~${\it T}$,~${\it P}$,~$R^{+}_{+}/R^{-}_{-}$~~~&~$\mathbb{Z}$~~\\[1ex]
		\hline\hline
	\end{tabular}
	\label{tab1}
\end{table}

\begin{figure*}[t]
\centering
\includegraphics[width=2.0\columnwidth]{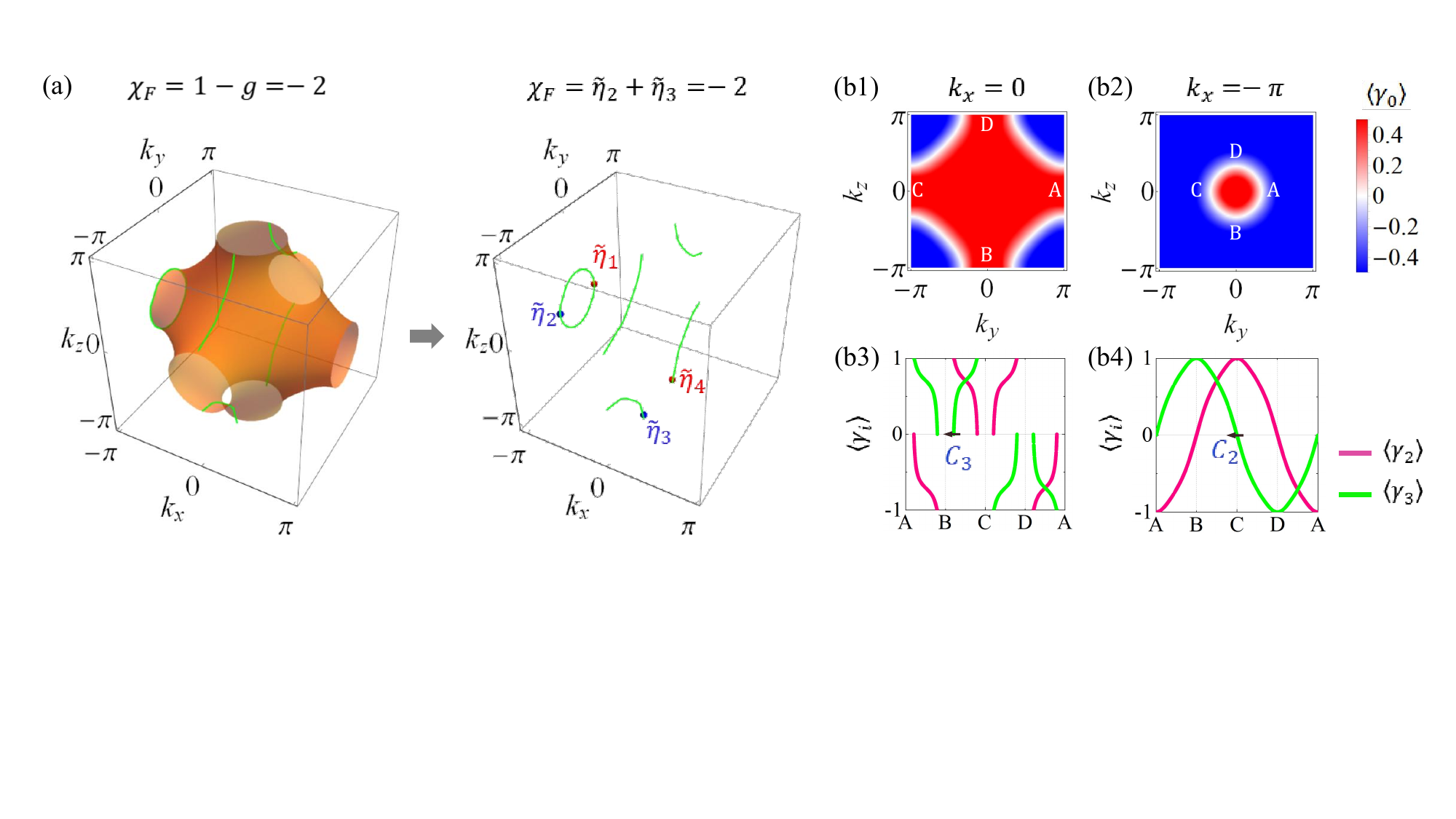}
\caption{(a) Reduction of Fermi sea topology in a 3D system. A 2D Fermi surface (orange color) with genus $g=3$ gives $\chi_F=1-g=-2$. Reducing it to obtain 1D curves (green color) on the Fermi surface, denoted as $\tilde{k}$, in which the reduced critical points with $\tilde{\eta}_2$ and $\tilde{\eta}_3$ determine $\chi_F=\tilde{\eta}_2+\tilde{\eta}_3=-2$. (b) Measuring $\chi_F$ of the normally filled band in a 3D topological superconductor. The pseudospin polarization of $\langle\gamma_{0,\mathbf{k}}\rangle$ at $k_x=0$ in (b1) and $k_x=-\pi$ in (b2) capture $\tilde{k}$ by $\langle\gamma_{0,\mathbf{k}}\rangle=0$. The topological charges ${\cal C}_{2,3}=-1$ are determined by $\langle\gamma_{3,\tilde{k}}\rangle=0$ and in the regions of $\langle\gamma_{2,\tilde{k}}\rangle>0$, showing in (b3) and (b4) and giving $\chi_F\Leftrightarrow{\cal C}_2+{\cal C}_3=-2$.  Here we have $\Delta_0=t_s$ and $\mu=-0.5t_s$.}
\label{fig:3}
\end{figure*}

For the $\mathbb{Z}$-classified topological SCs, we first confirm the above results using three typically models: the 1D Kitaev chain~\cite{kitaev2001unpaired}, the 2D $p-\mathtt{i}p$ SC~\cite{kallin2016chiral}, and the 3D He-3 B phase~\cite{leggett1975theoretical}. These models belong to the 1D class BDI, 2D class D, and 3D class DIII, respectively. In these models, the normal bands are described by $E_\mathbf{k}=-t_s\sum^d_{i=1}\cos k_i -\mu$ and the corresponding pairing order parameters take $\Delta_\mathbf{k}=(\Delta_0/t_s)\nabla_\mathbf{k}{E}_\mathbf{k}\cdot\mathbf{S}$. Here the matrices $\mathbf{S}$ are defined as $\mathbf{S}^{\text{BDI}}=-\mathtt{i}$ for the 1D class BDI, $\mathbf{S}^{\text{D}}=(1,-\mathtt{i})$ for the 2D class D, and $\mathbf{S}^{\text{DIII}}=(\sigma_z,-\mathtt{i}\sigma_0,\sigma_x)$ for the 3D class DIII. Explicitly, the pairing order parameters are given by 
\begin{equation}
\begin{split}
&\Delta^{\text{BDI}}_\mathbf{k}=-i\Delta_0\sin k_1,\\
&\Delta^{\text{D}}_\mathbf{k}=\Delta_0\sin k_1-i\Delta_0\sin k_2,\\
&\Delta^{\text{DIII}}_{\mathbf{k}}=\Delta_0(\sin k_1\sigma_z+\sin k_3 \sigma_x)-i\Delta_0\sin k_2.\\
\end{split}
\end{equation}
As shown in Fig.~\ref{fig:2}, the nontrivial result of $\chi_F=\mathcal{W}$ is confirmed, which implies that the number of Majorana edge states in these topological SCs can be exactly determined by $|\chi_F|$ of their normally filled bands.

Furthermore, we provide the general $\mathbb{Z}$-classified topological SCs satisfying Eq.~\eqref{pairing} in Tab.~\ref{tab1}, organized by the dimensionality $d=4n+j$, where $j=1,2,3,4$ and $n=0,1,2,\cdots$~\cite{kitaev2009periodic}. Inspired by the Ref.~\cite{ryu2010topological}, we define $2p+1$ anticommuting $2^p\times 2^p$ matrices from Clifford algebra as follows:
\begin{equation}
\begin{split}
&\Gamma^{2a-1}_{(2p+1)}=\underbrace{\sigma_0\otimes\cdots\sigma_0}_{a-1}\otimes\sigma_x\otimes\underbrace{\sigma_z\otimes\cdots\sigma_z}_{p-a},\\
&\Gamma^{2a}_{(2p+1)}=\underbrace{\sigma_0\otimes\cdots\sigma_0}_{a-1}\otimes\sigma_y\otimes\underbrace{\sigma_z\otimes\cdots\sigma_z}_{p-a},\\
&\Gamma^{2p+1}_{(2p+1)}=\sigma_z\otimes\sigma_z\cdots\otimes\sigma_z\otimes\sigma_z\otimes\cdots\sigma_z,\\
\end{split}
\end{equation}
where $a=1,2,\cdots,p$. The $(4n+j)$D $\mathbb{Z}$-classified topological SCs with odd (even) $n$ are obtained in the class BDI (CII) for $j=1$ or the class DIII (CI) for $j=2$ or the class D (C) for $j=3$. The corresponding BdG Hamiltonians are given by
\begin{equation}
\begin{split}
&\mathcal{H}^\text{BDI,CII,DIII,CI}_\mathbf{k}=\sum^{d=2p-1}_{\alpha=1}v_{\alpha,\mathbf{k}}\Gamma^{\alpha}_{(2p+1)}+E_\mathbf{k}\Gamma^{2p+1}_{(2p+1)},\\
&\mathcal{H}^\text{D,C}_\mathbf{k}=\sum^{d=2p}_{\alpha=1}v_{\alpha,\mathbf{k}}\Gamma^{\alpha}_{(2p+1)}+E_\mathbf{k}\Gamma^{2p+1}_{(2p+1)},
\end{split}
\end{equation}
with $E_\mathbf{k}=E_{-\mathbf{k}}$ and $v_{\alpha,\mathbf{k}}=-v_{\alpha,-\mathbf{k}}$. It should be noted that the $(4n+4)$D $\mathbb{Z}$-classified topological SCs need to be protected by the lattice symmetry. We consider the twofold-rotation symmetry with $R^{+}_{+}/R^{-}_{-}$ in the $\delta_\parallel=2$ family~\cite{shiozaki2014topology}, which belongs to the class C (D) with the even (odd) $n$. The corresponding BdG Hamiltonians are 
\begin{equation}
\begin{split}
&\mathcal{H}^\text{C}_\mathbf{k}=\sum^{d=2p}_{\alpha=1}v_{\alpha,\mathbf{k}}\Gamma^{\alpha}_{(2p+1)}+E_\mathbf{k}\Gamma^{2p+1}_{(2p+1)},\\
&\mathcal{H}^\text{D}_\mathbf{k}=\sum^{d=2p-2}_{\alpha=1}v_{\alpha,\mathbf{k}}\Gamma^{\alpha}_{(2p+1)}+\mathtt{i}E_\mathbf{k}\Gamma^{2p+1}_{(2p+1)}\Gamma^{2p}_{(2p+1)}\Gamma^{2p-1}_{(2p+1)},
\end{split}
\end{equation}
with $E_\mathbf{k}=-E_{-\mathbf{k}}$ and $v_{\alpha,\mathbf{k}}=v_{\alpha,-\mathbf{k}}$. It is observed that the topological invariant $\mathcal{W}$ of these topological (crystalline) SCs is equal to $\chi_F$ of their normally filled bands, when $\boldsymbol{\Gamma}$ have the same forms as $\boldsymbol{\gamma}$.

\section{Theoretical measurement of Fermi sea topology}
\label{measure}
In this section, we first show that the metallic systems can be induced as the topological SCs in Tab.~\ref{tab1} by introducing a Hubbard interaction into the $d$D filled band~\cite{liu2014realization,chan2017non,chan2017generic} or by adding a superconducting substrate to the metal sample to produce metal-superconductor heterostructure~\cite{eschrig2003theory,chung2011topological}. This implies that $\chi_F$ of these metallic systems can be determined through ${\cal W}$ of the corresponding topological SCs, where ${\cal W}$ is identified by using the (pseudo)spin polarizations $\langle\gamma_{i,\mathbf{k}}\rangle=\bra{u_\mathbf{k}}\gamma_i\ket{u_\mathbf{k}}$, with $i=0,1,\cdots,d$~\cite{loder2017momentum,gorol2022signatures,wang2023topological}. Here $\ket{u_\mathbf{k}}$ are the ground states of ${\cal H}_\mathbf{k}$. Specifically, the reduced critical points are determined by $\mathbf{k}_c \Leftrightarrow\left\{\mathbf{k}|\langle\gamma_{j,\mathbf{k}}\rangle=0; j=0,1,\cdots,d-2\right\}$ and then we have 
\begin{equation}
\tilde{\eta}_c=\frac{(-1)^q}{2}\left[\text{sgn}(\langle\gamma_{d,\mathbf{k}_{c,\text{L}}}\rangle)-\text{sgn}(\langle\gamma_{d,\mathbf{k}_{c,\text{R}}}\rangle)\right] 
\end{equation}
in the regions of $\langle\gamma_{d-1,\mathbf{k}}\rangle>0$. Here the right-hand side of the equation represents the measurements of topological charges in the topological SCs, which reflect the signatures of the critical points of the metallic systems on the left-hand side. Finally, we obtain $\chi_F$ as 
\begin{equation}
\chi_F=\sum_c\frac{{(-1)}^q}{2}\left[\text{sgn}(\langle\gamma_{d,\mathbf{k}_{c,\text{L}}}\rangle)-\text{sgn}(\langle\gamma_{d,\mathbf{k}_{c,\text{R}}}\rangle)\right].
\end{equation}
Compared with Eq.~\eqref{rcp}, this provided a scheme to measure $\chi_F$ using the (pseudo)spin polarizations, which may promote its experimental detection.

Next, a $3$D single filled band $E_\mathbf{k}=-t_s\sum^3_{i=1}\cos k_i-\mu$ is taken to demonstrate the dimensional reduction of $\chi_F$ and its measurements. First, we figure out the FS in Fig.~\ref{fig:3}(a) by setting $E_{\mathbf{k}}=0$ for $\mu=-0.5t_s$. It shows a 2D geometry with the genus $g=3$. Through the dimensional reduction, we obtain 1D momentum curves $\tilde{k}$ by setting $v_{1,\mathbf{k}}=0$ on the FS, which show the loop structures and are electron-like (hole-like) for $k_x=0$ ($k_x=-\pi$). We identify four reduced critical points where $v_{3,\tilde{k}}=0$, with the topological indices $\tilde{\eta}_{1,4}=1$ and $\tilde{\eta}_{2,3}=-1$. Here $\tilde{\eta}_{2,3}$ are located in regions of $v_{2,\tilde{k}}<0$ and give $\chi_F=\tilde{\eta}_2+\tilde{\eta}_3=-2$. Further, we show $\chi_F$ can be determined by the (pseudo)spin polarizations of the 3D topological SCs in class DIII [see Fig.~\ref{fig:2}(c)]. The corresponding BdG Hamiltonian is written as 
\begin{equation}
\begin{split}
{\cal H}_\mathbf{k}=&\left(-t_s\sum^3_{i=1}\cos k_i-\mu\right)\gamma_0+\Delta_0\sin k_x\gamma_1\\
&~~+\Delta_0\sin k_y\gamma_2+\Delta_0\sin k_z\gamma_3
\end{split}
\label{3D_SC}
\end{equation}
with $k_{1,2,3}=k_{x,y,z}$, where $\gamma_0=\tau_z$, $\gamma_1=\tau_x\sigma_z$, $\gamma_2=\tau_y$, and $\gamma_3=\tau_x\sigma_x$. Here $\boldsymbol{\sigma}$ and $\boldsymbol{\tau}$ are Pauli matrices acting on spin and Nambu degree of freedom, respectively. By calculating the expectation value of $\gamma_{0}$ at $k_x=0$ and $k_x=-\pi$, the 1D momentum curves $\tilde{k}$ are captured by $\langle\gamma_{0,\mathbf{k}}\rangle=0$, as shown in Figs.~\ref{fig:3}(b1) and \ref{fig:3}(b2). We further obtain $\langle\gamma_{2,\tilde{k}}\rangle$ and $\langle\gamma_{3,\tilde{k}}\rangle$ along $\tilde{k}$ which is in the clockwise direction, as shown in Figs.~\ref{fig:3}(b3) and \ref{fig:3}(b4). Four reduced critical points are determined by $\langle\gamma_{3,\tilde{k}}\rangle=0$. The topological charges in the regions of $\langle\gamma_{2,\tilde{k}}\rangle>0$ are identified by ${\cal C}_2=(\langle\gamma_{3,\tilde{k}_\text{R}}\rangle-\langle\gamma_{3,\tilde{k}_\text{L}}\rangle)/2=-1$ and ${\cal C}_3=(\langle\gamma_{3,\tilde{k}_\text{L}}\rangle-\langle\gamma_{3,\tilde{k}_\text{R}}\rangle)/2=-1$. And then, we immediately confirm
$\mathcal{C}_2=\tilde{\eta}_{2}$ and $\mathcal{C}_3=\tilde{\eta}_{3}$, and thus we have $\chi_F=\tilde{\eta}_{2}+\tilde{\eta}_{3}=-2$. Our scheme thus provides a feasible method to capture the Fermi sea topology, which may promote its experimental detection.  

\section{Discussion and Conclusion}
\label{discussion}

We have established a mapping between the Fermi sea topology of metals and the band topology of gapped systems. It is well-known that the band topology has exhibited a variety of intriguing phenomena, such as fractional charges~\cite{laughlin1981quantized}, fractional statistics~\cite{wilczek1982remarks}, and non-Abelian statistics~\cite{wen1991non}. This nontrivial mapping may provide a powerful framework to uncover the additional physical effects stemming from Fermi sea topology. Furthermore, we emphasize that the band topology can also be equivalently characterized through the Euler characteristic by defining a velocity field and applying the Poincar{\'e}-Hopf theorem~\cite{fan2021zero,fan2022topological}. However, this approach merely offers two equivalent topological descriptions of the same quantum topology. In contrast, our work reveals a profound connection between two distinct quantum topologies: Fermi sea topology and band topology, which are associated with the properties of the Fermi sea and the wave function, respectively.

In summary, we have developed a dimensional reduction theory for the Fermi sea topology in $d$-dimensional metallic systems, demonstrating that $\chi_F$ can be determined by the feature of reduced critical points on Fermi surfaces. This generic reduction theory updates the understanding of Fermi sea topology and provides a potential pathway for its experimental detection. Our work is expected to significantly advance the study of quantum topology in metals.

\begin{acknowledgments}
We thank Bao-Zong Wang, Yucheng Wang, and Yunhua Wang for helpful discussions. This work is supported by the National Natural Science Foundation of China (Grant No. 12404318 and No. 12247101), the Fundamental Research Funds for the Central Universities (Grant No. lzujbky-2024-jdzx06), the Natural Science Foundation of Gansu Province (No. 22JR5RA389), and the ‘111 Center’ under Grant No. B20063.
\end{acknowledgments}

\appendix

\section{Non-degenerate critical points}
\label{appendix-1}

For a NDCP, the Fermi velocity $\mathbf{v}_\mathbf{k}$ near $\mathbf{k}_m$ shows the linear dispersion. Hence we can simplify Eq.~\eqref{etm} as
\begin{equation}
\begin{split}
\eta_m=&\frac{\Gamma(d/2)}{2\pi^{d/2}}\int_{\mathcal{L}_m}\frac{J_{\mathbf{v}_\mathbf{k}}(\mathbf{k}_m)}{|\mathbf{v}_\mathbf{k}|^d}\sum^d_{i=1}(-1)^{i-1}k_{i}\bar{\bigwedge}^{d}_{s=1}\text{d}k_{s}\\
=&\text{sgn}[J_{\mathbf{v}_\mathbf{k}}(\mathbf{k}_m)]\frac{\Gamma(d/2)}{2\pi^{d/2}}\int_{\mathcal{L}_m}\frac{|J_{\mathbf{v}_\mathbf{k}}(\mathbf{k}_m)|}{|\mathbf{v}_\mathbf{k}|^d}\times\\
&\sum^d_{i=1}(-1)^{i-1}k_{i}\bar{\bigwedge}^{d}_{s=1}\text{d}k_{s}\\
=&\text{sgn}[J_{\mathbf{v}_\mathbf{k}}(\mathbf{k}_m)]\frac{\Gamma(d/2)}{2\pi^{d/2}}\int_{\mathbf{v}_\mathbf{k}(\mathcal{L}_m)}\sum^d_{i=1}(-1)^{i-1}\times\\
&\frac{v_{i,\mathbf{k}}}{|\mathbf{v}_\mathbf{k}|^d}\bar{\bigwedge}^{d}_{s=1}\text{d}v_{s,\mathbf{k}}\\
=&\text{sgn}[J_{\mathbf{v}_\mathbf{k}}(\mathbf{k}_m)],
\end{split}
\end{equation} 
where $J_{\mathbf{v}_\mathbf{k}}(\mathbf{k}_m)\equiv \text{det}\left[(\partial v_i/\partial k_j)|_{\mathbf{k}=\mathbf{k}_m}\right]$ denotes the Jacobian determinant and the integral gives the area of the $(d-1)$D sphere $\mathbf{v}_\mathbf{k}(\mathcal{L}_m)$, i.e., $\text{Vol}[\mathsf{S}^{d-1}]=2\pi^{d/2}/\Gamma(d/2)$.

\section{Dimensional reduction of band topology}\label{appendix-2}

The Ref.~\cite{yu2020high} has proposed a dimensional reduction theory for $d$D band topology, showing that the $d$D bulk topology for the gapped Hamiltonian
\begin{equation}
H_\mathbf{k}=v_{0,\mathbf{k}}\gamma_0+\sum^{d}_{i=1}v_{i,\mathbf{k}}\gamma_i
\label{ap_bulk} 
\end{equation}
can be reduced to a $(d-n)$D topological invariant defined in momentum subspace 
$\text{FS}^{(n-1)}\equiv\{\mathbf{k}|v_{0,\mathbf{k}}=v_{i_1,\mathbf{k}}=\cdots=v_{i_{n-1},\mathbf{k}}=0\}$, where $v_{i_\alpha,\mathbf{k}}$ are $n-1$ arbitrary components chosen as $v_{0,\mathbf{k}}$. And then, the remaining $d-n+1$ components construct an effective Fermi velocity field 
$\mathbf{v}^{(n-1)}_\mathbf{k}=(v_{i_n,\mathbf{k}},v_{i_{n+1},\mathbf{k}},\cdots,v_{i_d,\mathbf{k}})$ and this topological invariant is given by 
\begin{equation}
\begin{split}
\mathcal{W}=&\sum_l\frac{\Gamma[(d-n+1)/2]}{2\pi^{(d-n+1)/2}}\frac{1}{(d-n)!}\times\\
&~~~~~~~~~~~~~~~\int_{\text{FS}^{(n-1)}_l}\hat{\mathbf{v}}^{(n-1)}_\mathbf{k}\left[\text{d}\hat{\mathbf{v}}^{(n-1)}_\mathbf{k}\right]^{d-n},
\end{split}
\label{n_topo}
\end{equation} 
where $\hat{\mathbf{v}}^{(n-1)}_\mathbf{k}\equiv{\mathbf{v}^{(n-1)}_\mathbf{k}}/{|\mathbf{v}^{(n-1)}_\mathbf{k}}|$ denotes the unit effective Fermi velocity field. 

The above $\mathcal{W}$ is actually defined in the original BZ, which can be equivalently defined by a $(d-n+1)$D effective Hamiltonian 
\begin{equation}
H_{\tilde{\mathbf{k}}}=v_{i_{n-1},\tilde{\mathbf{k}}}\tilde{\gamma}_{i_{n-1}}+v_{i_{n},\tilde{\mathbf{k}}}\tilde{\gamma}_{i_{n}}+\cdots+v_{i_{d},\tilde{\mathbf{k}}}\tilde{\gamma}_{i_{d}},
\label{B3}
\end{equation}
where the effective Gamma matrices $\tilde{\boldsymbol{\gamma}}$ are defined in the effective BZ 
\begin{equation}
\tilde{\mathbf{k}}\equiv \{\mathbf{k}|v_{i_0,\mathbf{k}}=v_{i_1,\mathbf{k}}=\cdots=v_{i_{n-2},\mathbf{k}}=0\}.
\end{equation}
Here we have $i_\alpha\in\{0,1,\cdots,d\}$ with $\alpha=0,1,\cdots,d$. When further taking $\text{FS}^{(n-1)}\equiv\{\tilde{\mathbf{k}}|v_{i_{n-1},\tilde{\mathbf{k}}}=0\}$ and only once dimensional reduction is done, the effective Fermi velocity field is given by $\mathbf{v}^{(n-1)}_{\tilde{\mathbf{k}}}=(v_{i_n,\tilde{\mathbf{k}}},v_{i_{n+1},\tilde{\mathbf{k}}},\cdots,v_{i_d,\tilde{\mathbf{k}}})$. Correspondingly, $\mathcal{W}$ is rewritten as 
\begin{equation}
{\cal W}=\sum_{c\in {\cal V}_{\text{FS}^{(n-1)}}}{\cal C}_c,
\end{equation}
where the topological charge is simplified as 
\begin{equation}
\begin{split}
{\cal C}_c=&\frac{\Gamma[(d-n+1)/2]}{2\pi^{(d-n+1)/2}}\int_{\mathcal{L}_c}\frac{1}{|\mathbf{v}^{(n-1)}_{\tilde{\mathbf{k}}}|^{(d-n+1)}}\times\\
&~~~~~~~~~~~~\sum^{i_d}_{i=i_n}(-1)^{i-1}v_{i,\tilde{\mathbf{k}}}\bar{\bigwedge}^{d}_{s=n}\text{d}v_{s,\tilde{\mathbf{k}}}.
\end{split}
\end{equation}
Here ${\cal L}_c$ denotes a $(d-n)$D surface enclosing the $c$th topological charge at $\tilde{\mathbf{k}}=\tilde{\mathbf{k}}_c$ with $\mathbf{v}^{(n-1)}(\tilde{\mathbf{k}}_c)=0$. Here we have $\tilde{\mathbf{k}}_c=\{v_{i_n,\tilde{\mathbf{k}}}=v_{i_{n+1},\tilde{\mathbf{k}}}=\cdots=v_{i_d,\tilde{\mathbf{k}}}=0\}$, which is actually given in the original BZ by 
\begin{equation}
\begin{split}
\mathbf{k}_c&=\{v_{0,\mathbf{k}}=v_{i_\alpha,\mathbf{k}}=\cdots=v_{i_{n-2},\mathbf{k}}=\\
&~~~~~~~~~~~~v_{i_n,\mathbf{k}}=v_{i_{n+1},\mathbf{k}}=\cdots=v_{i_d,\mathbf{k}}=0\},
\end{split}
\end{equation}
where $v_{i_{n-1},\mathbf{k}}$ is non-zero and omitted.

\begin{figure}[t]
\centering
\includegraphics[width=1.0\columnwidth]{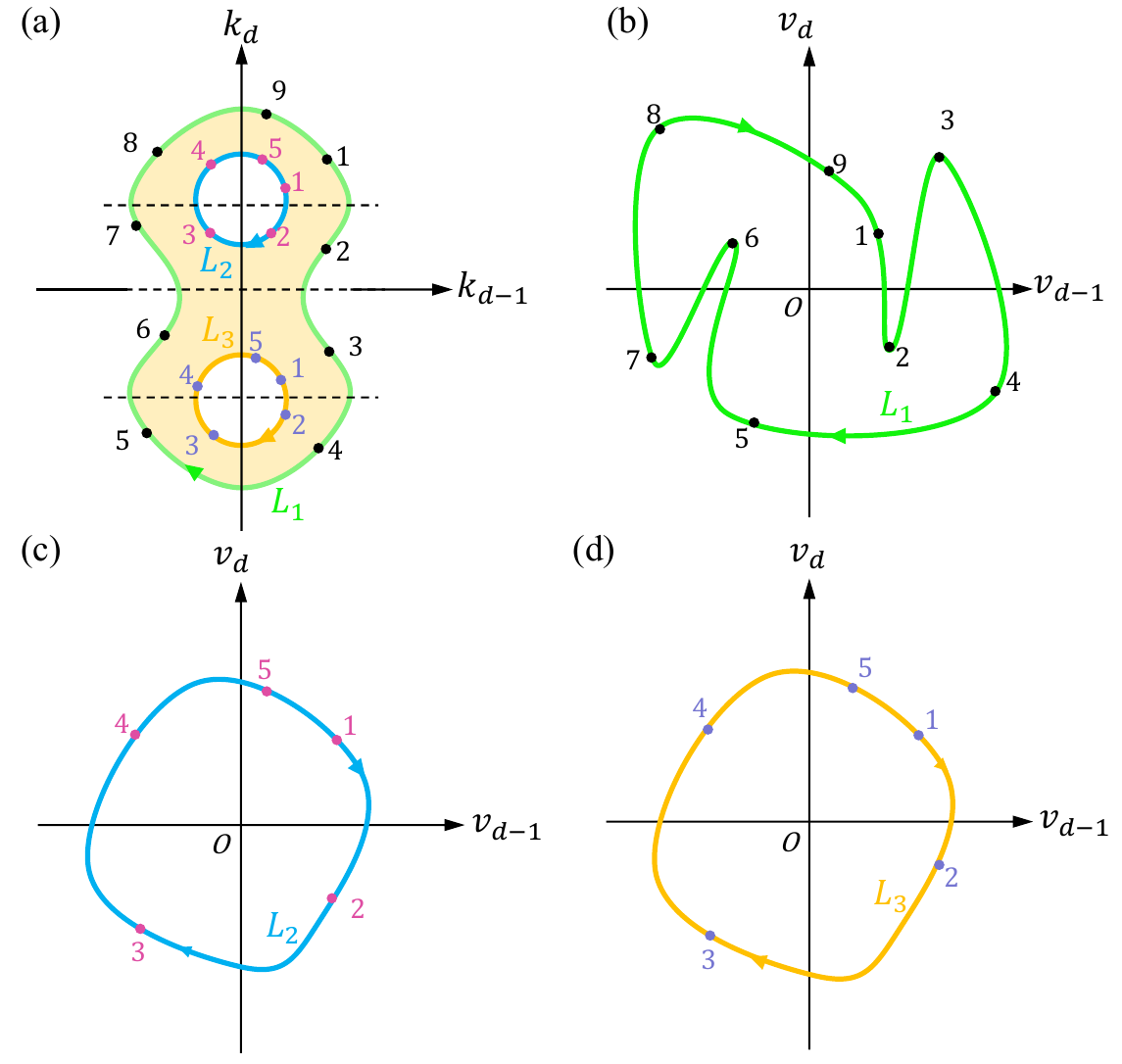}
\caption{Schematic diagram of choosing the direction of $\tilde{k}$. (a) The 1D effective $\text{FS}^{(d-2)}$ ($L_{1,2,3}$) in a 2D momentum subspace, where the electron-like (hole-like) $\tilde{k}$ is denoted as the green (lightblue or orange) curve, the black dashed lines give the locations of $v_d=0$, and the parameter points on $\text{FS}^{(d-2)}$ are marked by a set of serial number. (b)-(d) The direction of the integral path of the vector $(v_{d-1,\tilde{k}},v_{d,\tilde{k}})$ on the 1D $\text{FS}^{(d-2)}$ determines the direction of $\tilde{k}$, where $\tilde{k}$ in $L_{1,2,3}$ are clockwise.}
\label{fig:a1}
\end{figure}

\section{Dimensional reduction of $\chi_F$}\label{appendix-3}

Based on the above dimensional reduction of $\mathcal{W}$, we perform $d-1$ dimensional reduction, i.e. taking $n=d$ in Eq.~\eqref{B3}. As a result, the $d$D band topology is reduced to 1D winding number, which is defined by 1D effective Hamiltonian
\begin{equation}
H_{\tilde{k}}=v_{d-1,\tilde{k}}\tilde{\gamma}_{d-1}+v_{d,\tilde{k}}\tilde{\gamma}_{d}
\end{equation} 
with the effective Gamma matrices $\tilde{\gamma}_{d-1}$ and $\tilde{\gamma}_{d}$ defined in the $1$D effective momentum 
\begin{equation}
\tilde{k}\equiv \{\mathbf{k}|v_{0,\mathbf{k}}=v_{1,\mathbf{k}}=\cdots=v_{{d-2},\mathbf{k}}=0\}.
\end{equation}
Here we have $\text{FS}^{(d-1)}\equiv\{\tilde{k}|v_{d-1,\tilde{k}}=0\}$. Accordingly, the total topological charges defined by $H_{\tilde{k}}$ in the regions of $v_{d-1,\tilde{k}}<0$ [i.e., $\mathcal{V}_{\text{FS}^{(d-1)}}$] gives 
\begin{equation}
\mathcal{W}=\sum_c\mathcal{C}_c=\sum_c\text{sgn}\left(\frac{\partial v_{d,\tilde{k}}}{\partial \tilde{k}}\right),
\label{reducedcharge}
\end{equation}
of which these topological charges are located at 
$\tilde{k}_c\equiv\{\tilde{k}|v_{{d},\tilde{k}}=0\}
$. We further consider the mapping between the metallic systems and obtain
\begin{equation}
\tilde{\eta}_c=\frac{(-1)^q}{2}[\text{sgn}(v_{d,\tilde{k}_{c,\text{R}}})-\text{sgn}(v_{d,\tilde{k}_{c,\text{L}}})]
\end{equation}
where $q=0$ ($1$) is for the electron-like (hole-like) $\tilde{k}$. The subscripts ``R" and ``L" is the right- and left-hand point closest to $\tilde{k}_c$, respectively. Here the partial derivative in Eq.~\eqref{reducedcharge} has been written as a central difference in $\tilde{k}_c$ domain. It is noted that the direction of $\tilde{k}$ is determined by the integral path of the vector $\tilde{\mathbf{v}}'\equiv (v_{d-1,\tilde{k}},v_{d,\tilde{k}})$ on the 1D $\text{FS}^{(d-2)}$, as shown in Fig.~\ref{fig:a1}. Finally, the Euler characteristic is given by
\begin{equation}
\chi_F=\sum_c\frac{(-1)^q}{2}\left[\text{sgn}\left(v_{d,\tilde{k}_{c,\text{R}}}\right)-\text{sgn}\left(v_{d,\tilde{k}_{c,\text{L}}}\right)\right],
\label{Reduce_F_sm}
\end{equation}
which provides a simply characterization scheme for the Fermi sea topology. 

\begin{figure}[t]
\centering
\includegraphics[width=1.0\columnwidth]{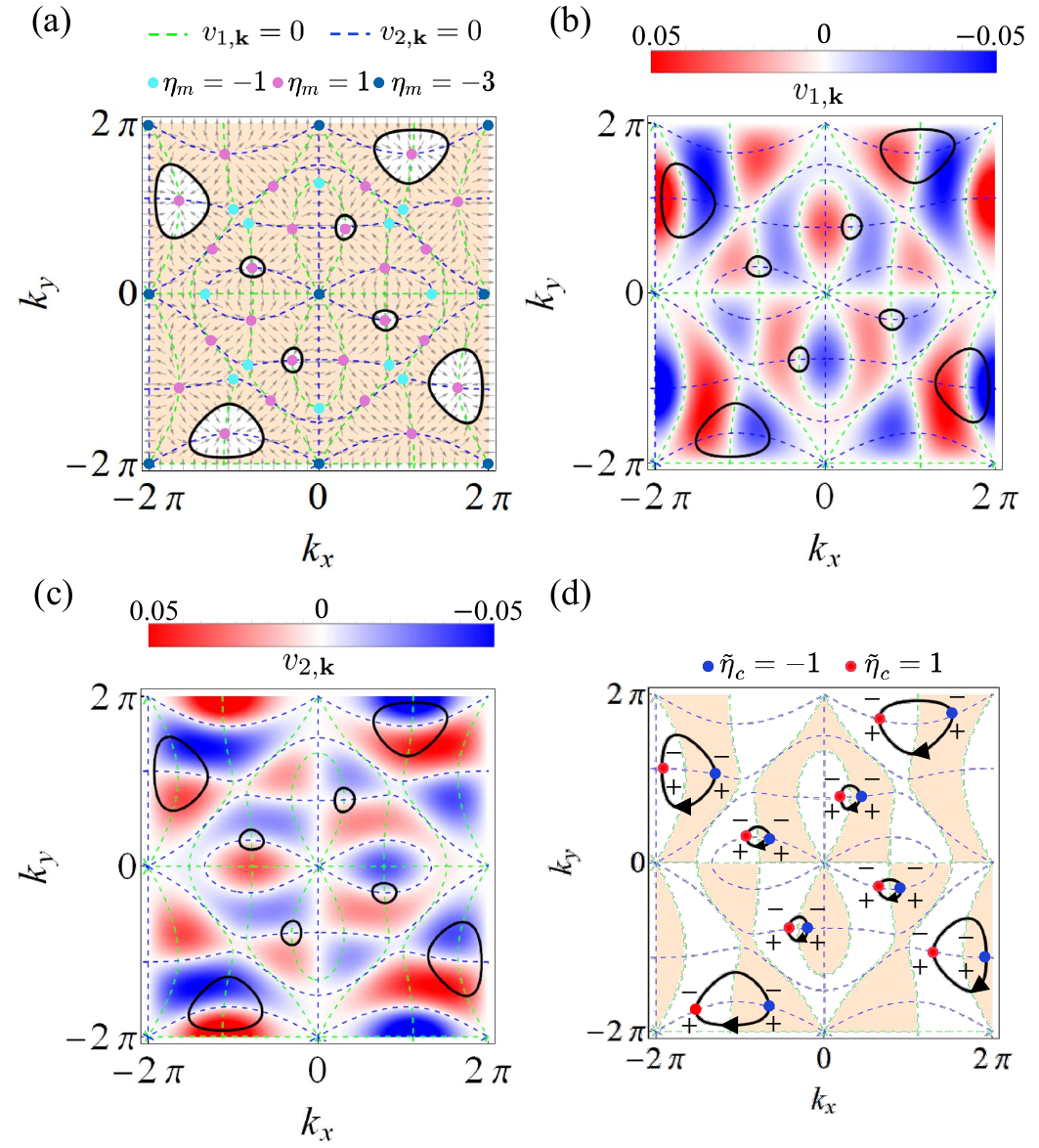}
\caption{(a) A 2D Fermi sea with eight hole-like FSs (black curves) in BZ and its spin texture of the unit vector field $\hat{\mathbf{v}}$, showing that only four critical points at $(k_x,k_y)=(0,0)$, $(0,-2\pi)$, $(-2\pi,0)$, and $(-2\pi,-2\pi)$ are DCPs with $\tilde{\eta_c}=-3$ and the others are NDCPs with $\tilde{\eta_c}=\pm 1$. Both $v_{1,\mathbf{k}}=0$ (green dashed lines) and $v_{2,\mathbf{k}}=0$ (blue dashed lines) determine forty critical points. There are thirty-two critical points in the filled band, and thus the Fermi sea topology is characterized by $\chi_F=\sum_m\eta_{m}=-8$. (b)-(c) The strength of $v_{1,\mathbf{k}}$ and $v_{2,\mathbf{k}}$. (d) Positive (red points) or negative (blue points) reduced critical points, located in 1D FSs, where $\tilde{k}_c$ in the region of $v_{1,\mathbf{k}_c}<0$ (orange regions) gives $\chi_F=\sum_c\tilde{\eta_c}=-8$. For a reduced critical point $\tilde{k}_c$ with $\tilde{\eta}_c=-1$, whose the nearest left (right) point $k_{c,\text{L}}$ ($k_{c,\text{R}}$) hosts an negative (positive) value for ($v_{2,k_{c,\text{L}}}$) $v_{2,k_{c,\text{R}}}$, marked as the green plus or minus. The black arrows give the direction of $\tilde{k}$. The similar case is shown for $\tilde{k}_c$ with $\tilde{\eta}_c=1$.}
\label{fig:a2}
\end{figure}

In order to show the advantages of our reduced characterization, we first use the traditional method to calculate $\chi_F$ and then compare the results with our theory. We employ the previous $2$D filled band $E_\mathbf{k}=e_{1,\mathbf{k}}e_{2,\mathbf{k}}e_{3,\mathbf{k}}-\mu$ with $e_{1,\mathbf{k}}=\sin^2(k_x/2)-\sin^2(k_y/2)$, $e_{2,\mathbf{k}}=\sin(k_x/2)\sin(k_y/2)$, and $e_{3,\mathbf{k}}=m_z-t_s\left[\cos(k_x/2)+\cos(k_y/2)\right]$. When choosing $m_z=0.5t_s$ and $\mu=0.18t_s$, we can see that $E_\mathbf{k}$ is partially filled and has eight hole-like FSs, as shown in Fig.~\ref{fig:a2}(a). Both $v_{1,\mathbf{k}}=0$ and $v_{2,\mathbf{k}}=0$ give 40 critical points in the BZ, of which four critical points at $(k_x,k_y)=(0,0)$, $(0,-2\pi)$, $(-2\pi,0)$, and $(-2\pi,-2\pi)$ are degenerate. We then define a unit vector field 
\begin{equation}
\hat{\mathbf{v}}\equiv\left({v_{1,\mathbf{k}}}{\sqrt{v^2_{1,\mathbf{k}}+v^2_{2,\mathbf{k}}}},{v_{2,\mathbf{k}}}{\sqrt{v^2_{1,\mathbf{k}}+v^2_{2,\mathbf{k}}}}\right) 
\end{equation}
to figure out the features of these critical points [see Fig.~\ref{fig:a2}(a)], where the winding of $\hat{\mathbf{v}}$ gives $\eta_m=-3$ to characterize these DCPs and the remaining critical points are non-degenerate with $\eta_m=\pm 1$. Hence we immediately identify that the Fermi sea topology is characterized by $\chi_F=\sum_m\eta_{m}=-8$, since there are 32 critical points in the filled band. Compared with the traditional method, we can plot the strength of $v_{1,\mathbf{k}}$ and $v_{2,\mathbf{k}}$ in Fig.~\ref{fig:a2}(b) and \ref{fig:a2}(c), respectively. By choosing $v_{1,\mathbf{k}}$ and performing the dimensional reduction for $\chi_F$, we observe that there are eight reduced critical points $\tilde{k}_c\equiv\{\mathbf{k}|E_\mathbf{k}=v_{2,\mathbf{k}}=0\}$ in the region of $v_{1,\tilde{k}}<0$, as shown in Fig.~\ref{fig:a2}(d). For the reduced critical point $\tilde{k}_c$ with $\tilde{\eta}_c=-1$, whose nearest left (right) point $k_{c,\text{L}}$ ($k_{c,\text{R}}$) hosts an negative (positive) value for ($v_{2,k_{c,\text{L}}}$) $v_{2,k_{c,\text{R}}}$. On the contrary, the reduced critical point with $\tilde{\eta}_c=1$, whose nearest left (right) point $k_{c,\text{L}}$ ($k_{c,\text{R}}$) hosts a positive (negative) value for $v_{2,k_{c,\text{L}}}$ ($v_{2,k_{c,\text{R}}}$). Since these FSs are hole-like and $q=1$, this determines the Fermi sea topology characterized by $\chi_F=\sum_c\tilde{\eta}_c=-8$.

\begin{figure*}[t]
\centering
\includegraphics[width=1.9\columnwidth]{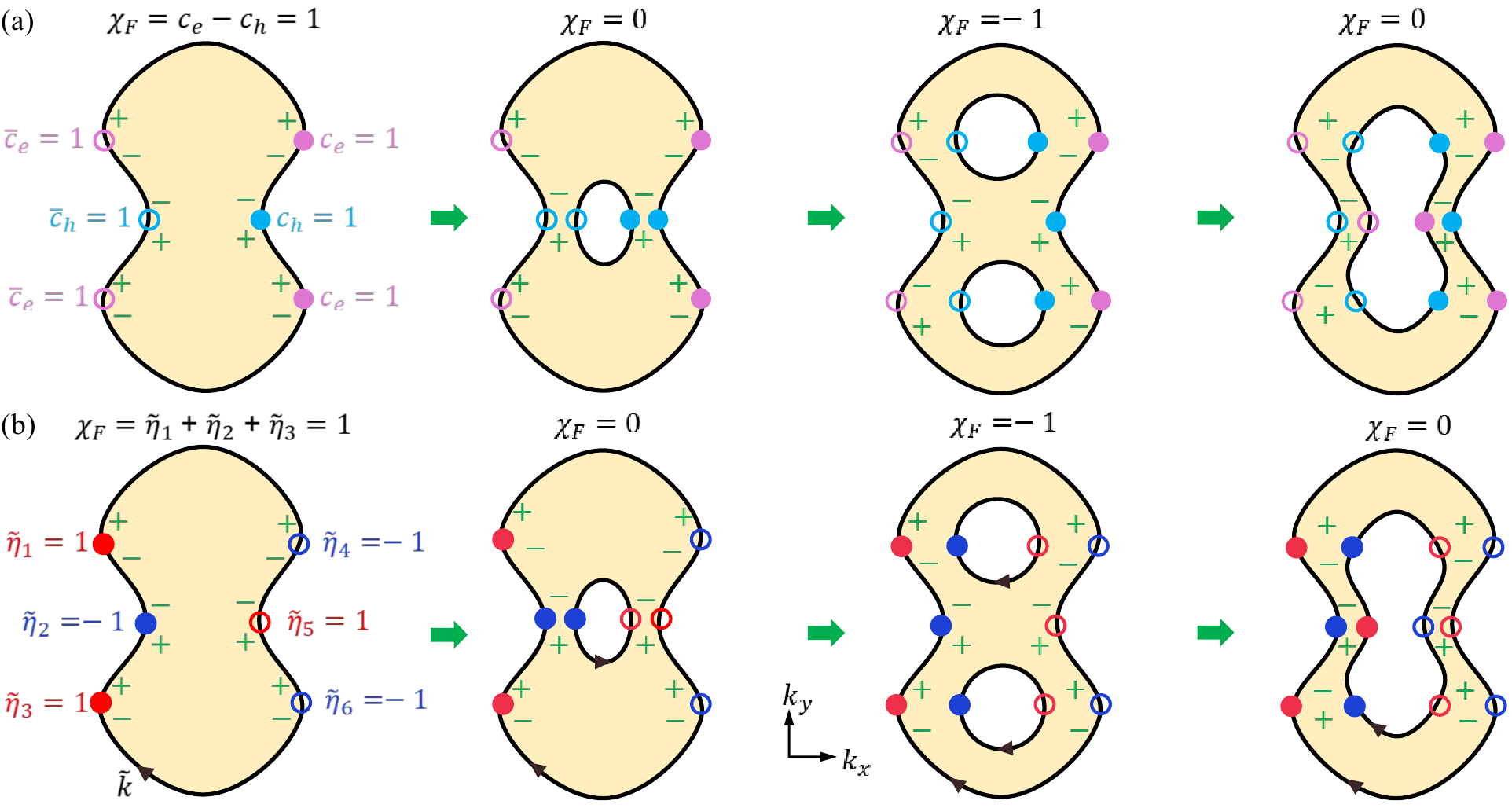}
\caption{Schematic diagram of 2D Fermi sea (orange regions), where the reduced critical points on the Fermi surfaces (black curves) determines the different $\chi_F$, driving the Lifshitz transitions $\chi_F=1\rightarrow 0 \rightarrow -1 \rightarrow 0$. (a) The convex/concave critical points, proposed in the Ref.~\cite{tam2023probing} and defined in $+k_x$ direction, are indicated by pink/lightblue dots with their count denoted as $c_{e/h}$. The Fermi sea topology is calculated by $\chi_F=c_e-c_h$. (b) The reduce critical points proposed in our theory are characterized by the zeroth Chern number and defined in $-k_x$ direction, which are indicated by red/blue dots with their count denoted as $\tilde{\eta}_{c}$. The direction of $\tilde{k}$ are marked by the black arrows and the Fermi sea topology is calculated by $\chi_F=\sum_c\tilde{\eta}_{c}$.}
\label{fig:a3}
\end{figure*}

\section{Reduced critical points in 2D systems}\label{appendix-4}

Here we consider the topological properties of the reduced critical points in 2D systems, which shows the fundamental difference with the previous studies~\cite{tam2023probing}. In Ref.~\cite{tam2023probing}, it has shown that $\mathbf{k}_c$ of the 2D metallic systems can capture the convexity (if $\partial v_{2,\mathbf{k}}/\partial k_{2}>0$) or concavity (if $\partial v_{2,\mathbf{k}}/\partial k_2<0$) of FSs. By denoting the number of convex or concave critical points as $c_{e/h}$, the Euler characteristic is given by $\chi_F=c_e-c_h$, where the convex (concave) critical points in the region of $v_{1}>0$ are only considered. These results are shown in Fig.~\ref{fig:a3}(a). For the case of $\chi_F=1$, it is seen that there are two convex and one concave critical points on the electron-like FS. With changing of the parameter, a new hole-like FS are emerged and on which one additional concave critical point is created. And then, the system hosts $\chi_F=0$. Moreover, $\chi_F$ can be further changed with increasing or decreasing the hole-like FSs. Hence the Lifshitz transitions are emerged in the changing of the filled Fermi seas, showing $\chi_F=1\rightarrow 0\rightarrow -1 \rightarrow 0$ in Fig.~\ref{fig:a3}(a).

In our theory, $\tilde{\eta}_c$ actually characterizes the zeroth Chern number of $\mathbf{k}_c$, which is determined along the $1$D FSs (i.e., the $1$D $\tilde{k}$). Following the rule of choosing direction of $\tilde{k}$ in Fig.~\ref{fig:a1}, we show the directions of $\tilde{k}$ in the above four different filled Fermi seas [see the black arrows in Fig.~\ref{fig:a3}(b)]. Since these FSs are all closed, the indexes ``L" and ``R" represents the previous and next closest point near the reduced critical point $\tilde{k}_c$, respectively. We show the results in Fig.~\ref{fig:a3}(b) and give $\chi_F=1\rightarrow 0\rightarrow -1 \rightarrow 0$ in the different Lifshitz transitions. It is clear that we have $\tilde{\eta}_{1,3}=1$ and $\tilde{\eta}_{2}=-1$ in the region of $v_{1}<0$ for the first case, with $\chi_F=\tilde{\eta}_{1}+\tilde{\eta}_{2}+\tilde{\eta}_{3}=1$. Compare with these two characterization schemes, it is seen that the convexity or concavity of $\mathbf{k}_c$ is determined by $\partial v_{2,\mathbf{k}}/\partial k_{2}\gtrless 0$, while the topological properties (i.e., the zeroth Chern number) of $\mathbf{k}_c$ is determined by $\partial v_{2,\tilde{k}}/\partial \tilde{k}\gtrless 0$, which are completely different [see the second and third cases in Fig.~\ref{fig:a3}(b), where the properties of $\mathbf{k}_c$ are different for the hole-like FS]. Thus our characterization of $\chi_F$ has fundamental difference with the Ref.~\cite{tam2023probing}. Remarkably, our theory provides an elegant and generic expression to identify $\chi_F$ via the reduced critical points on FSs, which covers the $d$D metallic bands. 

\section{Mapping the metallic band to topological superconductors}
\label{appendix-5}

\begin{figure*}[t]
\centering
\includegraphics[width=2.0\columnwidth]{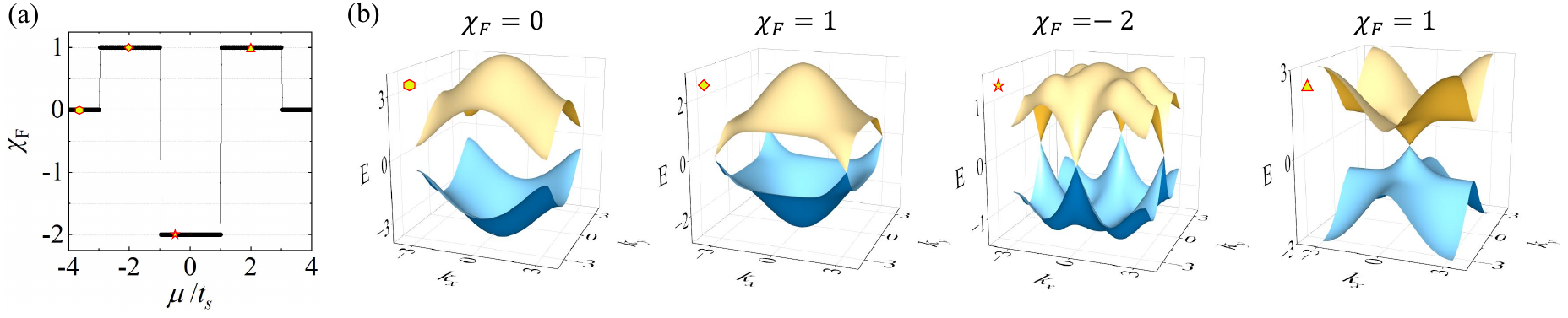}
\caption{(a) Numerical results of $\chi_F$ for the normal band of the 3D topological SCs. (b) The number of Majorana cone is given by $|\chi_F|$, where the parameters are $\mu=-3.5t_s$, $-2t_s$, $-0.5t_s$, and $2t_s$.}
\label{fig:a4}
\end{figure*}

We provide the complete classes of the topological SCs by mapping the single filled band. It is well-known that the ten-fold Altland-Zirnbauer classes~\cite{kitaev2009periodic} are defined through the presence or absence of time reversal symmetry ${T}$, particle-hole symmetry ${P}$, and chiral symmetry ${C}$, distinguishing the cases ${T}^2 = \pm 1$ and ${P}^2 = \pm 1$. Explicitly, the three symmetry operations read \begin{equation}
\begin{split}
  &\mathcal{H}(\mathbf{k}) =\, U_{T} \mathcal{H}(-\mathbf{k})^* U_{T}^{\dagger},\\
  &\mathcal{H}(\mathbf{k}) =\, - U_{P} \mathcal{H}(-\mathbf{k})^* U_{P}^{\dagger},\\
  &\mathcal{H}(\mathbf{k}) =\, - U_{C} \mathcal{H}(\mathbf{k}) U_{C}^{\dagger}, \\
\end{split}
\end{equation}
where $U_{T}$, $U_{P}$, and $U_{C}$ are $\mathbf{k}$-independent unitary matrices with $U_{T} U_{T}^* = {T}^2=\epsilon_T=\pm 1$ and $U_{P} U_{P}^* = {
P}^2=\epsilon_P=\pm 1$. If ${T}$ and ${P}$ are both present, one has $U_{C}=U_{P} U_{T}^*=T^2P^2U_TU_P^*$ and $U_C^2=C^2=1$. Besides, we need to consider the spatial symmetries $S$, which drives the systems to the topological crystalline SCs. Here we consider the order-two unitary symmetry [with $S$ being replaced by $M$ (mirror) $R$ (rotation), or $I$ (inversion) as need], which drives the system to satisfy 
\begin{equation}
U_S\mathcal{H}(\mathbf{k})U_S^{\dagger}=\mathcal{H}(-\mathbf{k}_\parallel,\mathbf{k}_{\perp}),
\end{equation}
with $U_SU^{*}_S=U_S^2=\epsilon_S=\pm 1$, where $\mathbf{k}_\parallel=-k_1,\cdots,-k_{d_\parallel}$ and $\mathbf{k}_{\perp}=k_{{d+1}_\parallel},\cdots,k_{d}$. It should be noted that $S$ can commute or anticommute with $T$, $P$, and $C$, i.e., $U_SU_T=\eta_T U_TU_S^*$, $U_SU_P=\eta_P U_PU_S^*$, and $U_SU_C=\eta_C U_CU_S$, where $\eta_T=\pm 1$, $\eta_P=\pm 1$, and $\eta_C=\pm 1$. Hence the order-two symmetries can be written as $S=S^{\epsilon_S}_{\eta_C}$, $S=S^{\epsilon_S}_{\eta_T}$, $S=S^{\epsilon_S}_{\eta_P}$, and $S=S^{\epsilon_S}_{\eta_T,\eta_P}$. 

First, the $\mathbb{Z}$-classified topological SCs can cover all dimensionality of the $d$D filled bands. For the $(4n+j)$D systems with $n=0,1,2,\cdots$, the $\mathbb{Z}$-classified TSCs are in the class BDI (CII) for $j=1$, or the class D (C) for $j=2$, or the class DIII (CI) for $j=3$, or the class C (D) for $j=4$, when $n$ is taken as even (odd). Thus we have 
\begin{equation}
\begin{split}
&\mathbb{Z}:~\text{BDI (1D)} \rightarrow \text{CII (5D)} \rightarrow \text{BDI (9D)},\\
&\mathbb{Z}:~\text{D (2D)} \rightarrow \text{C (6D)} \rightarrow \text{D (10D)},\\
&\mathbb{Z}:~\text{DIII (3D)} \rightarrow \text{CI (7D)} \rightarrow \text{DIII (11D)},\\
&\mathbb{Z}:~\text{C (4D)} \rightarrow \text{D (8D)} \rightarrow \text{C (12D)}.\\
\end{split}
\end{equation}
Note that there is no $\mathbb{Z}$-classified topological SCs for the $(4n+4)$D gapped systems with $n=0,1,2,\cdots$ under the above three intrinsic symmetries $T$, $P$, and $C$. Thus we hereby consider the twofold-rotation symmetry with $S=R$ and $d_\parallel=2$, which give these topological crystalline SCs in the class C or D where the systems have the particle-hole symmetry and host $S=S^{-}_{-}$ and $S^{+}_{+}$. 

Secondly, the $2\mathbb{Z}$-classified topological SCs for even (odd) $n$ can be covered in the class CII (BDI) for $j=1$, or the class C (D) for $j=2$, or the class CI (DIII) for $j=3$, or the class BDI (CII) for $j=4$ by the mapping of the metallic bands, and thus we have
\begin{equation}
\begin{split}
&2\mathbb{Z}:~\text{CII (1D)} \rightarrow \text{BDI (5D)} \rightarrow \text{CII (9D)},\\
&2\mathbb{Z}:~\text{C (2D)} \rightarrow \text{D (6D)} \rightarrow \text{C (10D)},\\
&2\mathbb{Z}:~\text{CI (3D)} \rightarrow \text{DIII (7D)} \rightarrow \text{CI (11D)},\\
&2\mathbb{Z}:~\text{BDI (4D)} \rightarrow \text{CII (8D)} \rightarrow \text{BDI (12D)}.\\
\end{split}
\end{equation}
Similarly, there is no $2\mathbb{Z}$-classified topological SCs for the $(4n+4)$D gapped systems with $n=0,1,2,\cdots$. Here we consider the order-two symmetry $S^+_{-+}/S^-_{+-}$, which give these topological crystalline SCs in the class BDI or CII. 

Finally, the $\mathbb{Z}_2^{(1)/(2)}$-classified topological SCs for even (odd) $n$ can also be covered by the mapping, which are in the class D/DIII (C/CI) for $j=1$, or the class DIII (CI) for $j=2$, or the class CII (BDI) for $j=3$, or the class CII/C (BDI/D) for $j=4$. Hence we have
\begin{equation}
\begin{split}
\mathbb{Z}_2^{(1)/(2)}&:~\text{D/DIII (1D)} \rightarrow \text{C/CI (5D)} \rightarrow \text{D/DIII (9D)},\\
\mathbb{Z}_2^{(1)}&:~\text{DIII (2D)} \rightarrow \text{CI (6D)} \rightarrow \text{DIII (10D)},\\
\mathbb{Z}_2^{(2)}&:~\text{CII (3D)} \rightarrow \text{BDI (7D)} \rightarrow \text{CII (11D)},\\
\mathbb{Z}_2^{(1)/(2)}&:~\text{CII/C (4D)} \rightarrow \text{BDI/D (8D)} \rightarrow \text{CII/C (12D)}.\\
\end{split}
\end{equation}

With this we provide a large class of topological SCs, where $\chi_F$ of their normally filled bands are exactly equal to their topological invariant ${\cal W}$, with which the $\mathbb{Z}$-, $2\mathbb{Z}$-, and $\mathbb{Z}_2$-classified topological SCs are characterized by $\mathcal{W}$, $2\mathcal{W}$, and $(-1)^\mathcal{W}$, respectively. Accordingly, the number of Majorana edge modes is determined by $|\chi_F|$ of their normally filled bands. As shown in Fig.~\ref{fig:a4}, we numerically calculate the $\mathbb{Z}$-classified 3D topological SC described by Eq.~\eqref{3D_SC}, showing that the number of Majorana cone is exactly equal to $|\chi_F|$.

\bibliography{references}
\bibliographystyle{apsrev4-1}
\end{document}